\documentclass[aps,twocolumn,showpacs,preprintnumbers,amssymb]{revtex4}

\newcommand{\be}{\begin{eqnarray}}
\newcommand{\ee}{\end{eqnarray}}

\newcommand{\eins}{\mbox{$1 \hspace{-1.0mm}  {\bf l}$}}
\def\bea{\begin{eqnarray}}
\def\eea{\end{eqnarray}}
\def\C{\hbox{$\mit I$\kern-.7em$\mit C$}}
\def\N{\hbox{$\mit I$\kern-.3em$\mit N$}}

\usepackage{dcolumn} % Align table columns on decimal point
\usepackage{bm} % bold math

\usepackage{epsfig}

\begin{document}
 %\draft

\title{Non--additivity of quantum capacity for multiparty communication channels}

\author{W. D\"{u}r$^{1}$, J. I. Cirac$^2$ and P. Horodecki$^3$}

\affiliation{
$^1$Institut f\"ur Theoretische Physik, Universit\"at Innsbruck, A-6020 Innsbruck, Austria.\\
$^2$Max--Planck Institut f\"ur Quantenoptik, Hans--Kopfermann Str. 1, D-85748 Garching, Germany.\\
$^3$Faculty of Applied Physics and Mathematics, Gda{\'n}sk University of Technology, 80-952 
Gda{\'n}sk, Poland.}

\date{\today}

\begin{abstract}
We investigate multiparty communication scenarios where information is sent 
from several sender to several receivers. We establish a relation between the 
quantum capacity of multiparty communication channels and their distillability 
properties which enables us to show that the quantum capacity of such channels 
is not additive.
 \end{abstract}

\maketitle
 %\bigskip

 %-------------------------------------------------------------------
 % Introduction
 %-------------------------------------------------------------------

The emerging field of quantum information theory is a fruitful combination of quantum 
theory and classical information theory, leading to surprising new insights into both fields.
 One of the most important goals in the development of a quantum information theory is to 
provide analogs of the central theorems of classical information theory, most notable 
Shannon's noiseless and noisy coding theorem \cite{Sh48}. For noiseless quantum channels 
such an analog has been found by Schumacher \cite{Sc95,Jo94}.
The analog theorem known for noisy quantum channels
\cite{Shor03} exploiting coherent information does not provide 
a closed analytical formula for the capacity of a given noisy quantum channel.
Finding the latter is, in general, a difficult problem because 
it involves maximization over all possible coding and decoding procedures. 

One of the major open problems in this context is the question 
whether quantum channel capacities are {\it additive}.
Following pioneering papers connecting channel capacities 
with entanglement distillation \cite{Be97}
and identifying bound entanglement \cite{bound}, it has been conjectured \cite{bechan,APP}
that channels capacities are not additive, i.e. that there exist
quantum channels ${\cal E}_j$, $j=1,2$
such that their capacities $Q({\cal E}_j)$ 
satisfy a {\it superadditivity} relation,
\be
Q({\cal E}_1\otimes{\cal E}_2) > Q({\cal E}_1)+Q({\cal E}_2).\label{addi}
\ee 
Obtaining equality in this expression for all channels ${\cal E}_j$ would on the 
other hand imply additivity of the quantum channel capacity. Despite of considerable 
effort, this question remains unanswered so far. 

One of the candidates for nonadditivity (\ref{addi})
in the bipartite scenario (one sender and one receiver)
are so called {\it binding entanglement (BE) channels} introduced in 
\cite{bechan,CMP}, i.e.  channels that can produce only bound entanglement 
\cite{bound} after sending one of any given two entangled subsystems. 
BE channels have all capacities zero \cite{bechan,CEJP} and their 
possible nonadditivity is connected to the 
conjectured existence of NPT bound entanglement \cite{APP}.
%``NPT bound entanglement conjecture'' \cite{APP}.

In this paper, we consider multiparty communication scenarios where quantum information 
is sent from several senders to several receivers through a noisy quantum channel. 
For such multiparty communication channels, natural generalizations of the definition 
of quantum channel capacities are possible. 
We establish a connection of different kinds of quantum channel capacities to the 
capability of the channel to create (distillable) multipartite entangled states. 
We show for certain scenarios that the ability of a channel to faithfully transmit 
quantum information is equivalent to its capability to generate a certain kind of 
(distillable) multipartite entangled states. This connection allows us to show that 
all these multipartite quantum channel capacities are, in general, not additive. 
The considered channels are multipartite versions of {\it binding entanglement channels}.
We give an example of several such channels 
all of which have zero capacity (and cannot produce pure entangled states), 
while simultaneous availability of all these channels allows to 
faithfully transmit quantum information, 
i.e. the new channel has non--zero capacity.
% (and can produce arbitrary multipartite entangled pure states). 

 %-------------------------------------------------------------------
 % Quantum channel capacity
 %-------------------------------------------------------------------

{\bf 1. Quantum channel capacity}\\
We consider a quantum channel ${\cal E}$ which is described by a completely positive 
trace preserving linear map ${\cal E}: B({\cal H}_c) \to B({\cal H}_o)$ from the space 
$B({\cal H}_c)$ of bounded linear operators on the input Hilbert space ${\cal H}_c$ 
to the space $B({\cal H}_o)$ of bounded linear operators on the output Hilbert space 
${\cal H}_o$. We consider isomorphic system Hilbert spaces ${\cal H}_{s_{in}}$ and 
${\cal H}_{s_{out}}$ and coding (${\cal C}$) and decoding (${\cal D}$) operations, 
where ${\cal C}: B({\cal H}_{s_{in}}) \to B({\cal H}_c^{\otimes n})$ and 
${\cal D}: B({\cal H}_o^{\otimes n}) \to B({\cal H}_{s_{out}})$. These coding and 
decoding operations define a $(n,\epsilon)$ code if
\be
{\rm min}_{|\phi\rangle \in{\cal H}_{s_{in}}}F(|\phi\rangle,({\cal D} 
\circ {\cal E}^{\otimes n} \circ {\cal C}) |\phi\rangle\langle\phi|) \geq 1-\epsilon.
\ee
We remark that we implicitly made use of the isomorphy between ${\cal H}_{s_{in}}$ 
and ${\cal H}_{s_{out}}$ to define the fidelity $F$ between two states on different 
Hilbert spaces, that is for some arbitrary but fixed bases $\{|\phi_i\rangle\}$ and 
$\{|\chi_i\rangle\}$ with $|\phi_i\rangle \in {\cal H}_{s_{in}}$ and $|\chi_i\rangle \in 
{\cal H}_{s_{out}}$, we identify $|\phi_i\rangle \equiv |\chi_i\rangle$. The rate 
$R\equiv 1/n \log \dim {\cal H}_{s_{in}}$ of the code is called achievable if for all 
$\epsilon,\delta >0$ and sufficiently large $n$ there exists a code of rate $R-\delta$. 
The quantum capacity $Q(\cal E)$ of a bipartite quantum channel ${\cal E}$ is defined as
 the supremum of all achievable rates $R$ (see Ref. \cite{Ba98} for a rigorous definition).
 Coding and decoding operations may be assisted by forward classical communication 
($\rightarrow$) or two--way classical communication $(\leftrightarrow)$ which gives 
rise to quantum capacities $Q^{\rightarrow}$ [$Q^{\leftrightarrow}$] respectively. 
We remark that a minimal pure state fidelity $F=1-\epsilon$ for all $|\phi\rangle \in 
{\cal H}_{s_{in}}$ implies an entanglement fidelity $F_e \geq 1- 3/2\epsilon$ for all 
density operators $\rho$ whose support lies entirely in that subspace \cite{Ba98}. 
That is, when transmitting part of an entangled state $|\Psi\rangle$ which is a 
purification of $\rho$, we have that the resulting state has fidelity $F\geq 1-3/2 
\epsilon$ with respect to $|\Psi\rangle$.

% Multiparty quantum channels: N senders and M receivers

\begin{figure}[ht]
\begin{picture}(230,90)
\put(-5,-5){\epsfxsize=230pt\epsffile[7 690 330 833]{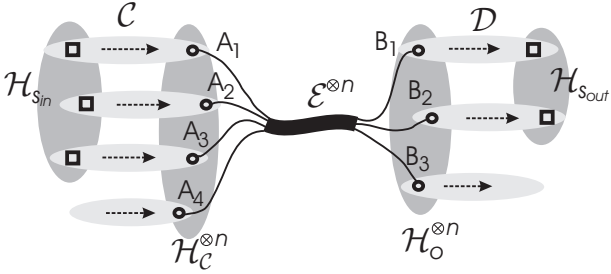}}
\end{picture}
\caption[]{Multiparty communication scenario with ${\tilde A}=A_1 A_2 A_3, {\tilde B}=B_1B_2$. Note that party $A_4 \not\in \tilde A$ may prepare its system in some suitable state, while party $B_3 \not\in \tilde B$ can e.g. perform suitable measurements to assist the communication.}
\label{setup}
\end{figure}

One can generalize the definition of channel capacity to a multipartite scenario with 
$N$ spatially separated senders $A_1, \ldots ,A_N$ and $M$ spatially separated receivers 
$B_1,\ldots ,B_M$. In this case, the input Hilbert space is given by the tensor product 
of the Hilbert spaces of the senders, ${\cal H}_c = {\cal H}_{A_1} \otimes \ldots 
\otimes{\cal H}_{A_N} $, and the output Hilbert space is a tensor product of the Hilbert 
spaces of the receivers, ${\cal H}_o={\cal H}_{B_1}\otimes \ldots \otimes {\cal H}_{B_M}$,
 while the quantum channel ${\cal E}: B({\cal H}_c) \to B({\cal H}_o)$ as before. In such 
a scenario the allowed operations (in particular ${\cal C}$ and ${\cal D}$) are restricted 
to {\em local} operations $A_1-\dots -A_N - B_1-\ldots -B_M$. We denote by LOCC$^{C}$ local 
operations assisted some kind $C$ of classical communication between senders and receivers 
(e.g. $C=\leftrightarrow$ denotes two way classical communication between each sender $A_i$ 
and all receivers $B_j$). For any subset of sender ${\tilde A}$ and any subset of receivers 
${\tilde B}$ one can define a channel capacity $Q_{\tilde A \to \tilde B}$ which measures 
the amount of quantum information that can be sent from the set of senders $\tilde A$ to 
the set of receivers $\tilde B$. In that case, we have that ${\cal H}_{s_{in}} = 
{\cal H}_{\tilde A} \cong  {\cal H}_{s_{out}} = {\cal H}_{\tilde B}$ and all operations 
are LOCC$^{C}$, where ${\cal H}_{\tilde A}=\bigotimes_{A_j \in \tilde A}\tilde{\cal H}_{A_j}$ is some (arbitrary) Hilbert space of parties $A_j \in \tilde A$ and similarly for ${\cal H}_{\tilde B}$ (see Fig. \ref{setup}).

We point out that although $Q_{\tilde A \rightarrow \tilde B} >0$, one might have the 
situations that (i) the parties $A_k \in \tilde A$ cannot transmit information if they 
prepare the states of their systems separately and (ii) the information is distributed 
among the parties $B_j \in \tilde B$ and cannot be accessed locally. Situation (i) might 
appear when a channel is only capable to transmit entangled states, while (ii) might occur
 when the output states of a channel are entangled. 
If some senders $A_k$ and some receivers $B_j$ are not involved in the transmission, 
and are excluded from classical communication with the remaining system, one can 
equivalently consider the reduced channel $\tilde{\cal E}$ which corresponds to the 
initial channel ${\cal E}$ followed by trace over all parties $A_k$,$B_j$.

Given the variety of possible choices of $\tilde A$, $\tilde B$ and the allowed classical
 communication $C$, %among the parties
it is not hard to imagine that such a multiparty 
communication system is very rich and displays many new 
and interesting features \cite{Notefeatures}.

 %-------------------------------------------------------------------
 % Multiparty distillability
 %-------------------------------------------------------------------
{\bf 2. Distillability in multipartite systems}\\
In the following, we establish a connection of the different channel capacities to 
different kinds of multipartite distillability. Consider again a multi--local scenario
 with spatially separated parties $\{A_k\}$ and $\{B_j\}$ and subsets $\tilde A$, 
$\tilde B$ and some kind of classical communication $C$. We have that $\rho_{AB}$ 
is distillable in the sense $\tilde A\tilde B$ if an entangled state shared between
 the systems ${\tilde A}$ and ${\tilde B}$ can be created from several copies of 
$\rho_{AB}$ by means of LOCC$^C$. That is, a state $\rho_{AB}$ has 
$D_{\tilde A\tilde B}^C(\rho) >0$ if and only if there exists a (multilocal) 
transformation ${\cal E}: B({\cal H}_{AB}^{\otimes N}) \to B({\cal H}_{\tilde A}
\otimes{\cal H}_{\tilde B})$ such that $\rho_{AB}^{\otimes N} \rightarrow_{{\rm LOCC}^C} 
\sigma_{\tilde A\tilde B}$ with $\langle \Phi|\sigma|\Phi\rangle \geq 1-\epsilon$ 
$\forall \epsilon >0$, where $|\Phi\rangle \equiv 1/\sqrt{2}(|\chi_1\rangle_{\tilde A} 
|\psi_1\rangle_{\tilde B}+ |\chi_2\rangle_{\tilde A} |\psi_2\rangle_{\tilde B})$ and 
$|\chi_1\rangle,|\chi_2\rangle$ [$|\psi_1\rangle,|\psi_2\rangle$] are some orthogonal 
states of the system ${\tilde A}$ [${\tilde B}$] respectively. We remark that the 
states $|\chi_k\rangle, |\psi_j\rangle$ might itself be entangled, which can imply 
that $|\Phi\rangle$ cannot be used to accomplish certain tasks (e.g. teleportation)
 by means of only local operations. 
%if classical communication $C$ are restricted.

We proceed by pointing out some non--trivial relations between different kinds of 
distillability. First we have that $\exists B_j\in \tilde B$, 
$D_{\tilde A B_j}^C(\rho_{AB})>0 \Rightarrow D_{\tilde A \tilde B}^C(\rho_{AB})>0$. 
The converse is, however, not generally true. Consider for instance a tripartite GHZ 
state, $|\Psi\rangle_{AB_1B_2} =1/\sqrt{2}(|000\rangle+|111\rangle)$ and LOCC$^\rightarrow$.
 Clearly, $D_{A (B_1B_2)}^\rightarrow(|\Psi\rangle) >0$, however $D_{AB_j}^\rightarrow
(|\Psi\rangle)=0 \forall j$. Note that e.g. the state of the system $AB_1$ is described 
by the reduced density operator $\rho_{AB_1}=1/2(|00\rangle\langle 00|+|11\rangle\langle 11|)$
 regardless of the operation performed at $B_2$, since the results of possible 
measurements in $B_2$ cannot be communicated to $A$ or $B_1$. We have that $\rho_{AB_1}$ 
is separable and hence $D_{AB_1}(|\Psi\rangle)=0$.

On the other hand for LOCC$^{\leftrightarrow}$ and $A_k\in \tilde A$, $B_j\in \tilde B$ 
one can show that
\be
\exists (A_k,B_j), D_{A_kB_j}^\leftrightarrow(\rho_{AB})>0 \Leftrightarrow D_{\tilde A\tilde B}^\leftrightarrow(\rho_{AB})>0,\label{equivD}
\ee
that is the possibility to create entanglement between the composed systems $\tilde A$ 
and $\tilde B$ is equivalent to the possibility to create entanglement between at 
least one of the individual parties $A_k$ and $B_j$. It remains to show that from 
$|\Phi\rangle \equiv 1/\sqrt{2}(|\chi_1\rangle_{\tilde A} |\psi_1\rangle_{\tilde B}+ 
|\chi_2\rangle_{\tilde A} |\psi_2\rangle_{\tilde B})$ with $|\chi_1\rangle \not= 
|\chi_2\rangle$, $|\psi_1\rangle \not= |\psi_2\rangle$ one can create a maximally 
entangled state shared between $A_k$ and some $B_j$ by means of LOCC$^{\leftrightarrow}$.

This can be seen using the following lemma (see also \cite{Po92}): For all states $|\Psi_0\rangle_{A_1\ldots
 A_M} \not= |\Psi_1\rangle_{A_1\ldots A_M}$ which are not of the form $(i)$ $|\Psi_0\rangle =
|\varphi_0\rangle_{A_1} \otimes |\Phi_0\rangle_{A_2\ldots A_M}$ and $|\Psi_1\rangle=
|\varphi_1\rangle_{A_1}\otimes |\Phi_0\rangle_{A_2\ldots A_M}$ with $|\varphi_0\rangle 
\not= |\varphi_1\rangle$), there exists a projector $P_{A_1} \equiv|\varphi\rangle_{A_1}
\langle\varphi|$ such that the resulting states $|\tilde\Psi_j\rangle \equiv P_{A_1}
|\Psi_j\rangle/||\tilde\Psi_j\rangle||$ fulfill $|\tilde\Psi_0\rangle \not= 
|\tilde\Psi_1\rangle$ . 
The proof is by contradiction: Assume on the opposite that $|\tilde\Psi_0\rangle = 
|\tilde\Psi_1\rangle$ for all projectors $P_{B_1}$. One readily convinces oneself that
 this implies that either $|\Psi_0\rangle = |\Psi_1\rangle$ or $(i)$ is fulfilled, from 
which the lemma follows. 

We sequentially apply the lemma to systems $A_1,A_2,\ldots A_{M-1}$ and stop if (i) 
applies at some point. In step one we have e.g. that either $|\chi_1\rangle,|\chi_2\rangle$ 
fulfills $(i)$, which leaves parties $A_1$, $\tilde B$ with a state of the form 
$(|\varphi_1\rangle_{A_1}|\psi_1\rangle_{\tilde B} + |\varphi_2\rangle_{A_1}|\psi_2
\rangle_{\tilde B})$ while the other parties are factored out. If $(i)$ does not apply, 
then from the lemma follows that there exists a projective measurement in $A_1$ such 
that the resulting state of systems $A_2,\ldots,A_M\tilde B$ is of the same form as the 
initial state $|\Phi\rangle$, but the number of parties $A_k$ is decreased by one 
(and party $A_1$ is factored out). Doing the same in system $\tilde B$, we have that 
one ends up with a state of the form $(|\varphi_1\rangle_{A_k}|\gamma_1\rangle_{B_j} + 
|\varphi_2\rangle_{A_k}|\gamma_2\rangle_{B_j})$ shared between two parties $A_k$ and 
$B_j$ for some $(k,j)$, while all other parties are factored out. This state is 
distillable, e.g. by means of filtering measurements in $A_k$ and $B_j$ which ends
 the proof of Eq. (\ref{equivD}).

 %-------------------------------------------------------------------------------------

{\bf 3. Relation between $Q_{A\rightarrow \tilde B}$ and $D_{A\tilde B}$}\\
We will now show the qualitative equivalence of non--zero multipartite channel 
capacities and the capability of the channel to create distillable entanglement. 
We will restrict ourselves to channels with only a {\em single} sender $A\equiv A_1$ 
and several receivers $B_1,\ldots, B_M$ and classical communication $C$ which contains
 at least forward communication from $A$ to all receivers $B_j\in\tilde B$. Our results 
are a generalization of the results found in Ref. \cite{CEJP} for bipartite communication 
channels to the multipartite case. 

A nonzero channel capacity $Q^{C}_{A\rightarrow \tilde B}$ implies --by definition-- that
 there exists a LOCC$^{C}$ implementable 
coding and decoding procedure such that a subspace ${\cal H}$ with 
$\dim {\cal H} \geq 2$ can be reliable transmitted. Since the entanglement 
fidelity is $F_e = 1-3/2 \epsilon$ with $\epsilon$ arbitrarily small, one can 
use this coding and decoding procedure to successfully transmit one half of a 
maximally entangled state of two qubits $|\Phi\rangle = 1/\sqrt{2}
|0\rangle_A|\Psi_0\rangle + |1\rangle|\Psi_1\rangle)$ with $\langle \Psi_0|\Psi_1\rangle=0$,
 $|\Psi_j\rangle \in {\cal H}_{s_{in}} \cong {\cal H}_{\tilde B}$ such that the resulting 
state $\rho_{A\tilde B}$ fulfills $F=\langle \Phi|\rho_{A\tilde B}|\Phi\rangle \geq 1-3/2 
\epsilon$. This already shows that a maximally entangled state shared between $A$ and the 
joint system $\tilde B$ can be created by means of LOCC$^C$ if $Q^C_{A\rightarrow 
\tilde B}>0$. In turn we have that whenever such a state $|\Phi\rangle_{AB}$ 
is available it can be used to faithfully transmit quantum information by means
 of teleportation (which only involves forward classical communication) \cite{Be97}. 
Note, however, that the operations required in the original teleportation scheme to
 achieve deterministic teleportation for all possible measurement outcomes may not 
be locally implementable since the states $|\Psi_j\rangle$ are not necessarily of 
product form. As we did not fix a basis in ${\cal H}_{s_{out}}$ in the definition of
 $Q_{A\rightarrow \tilde B}$, we have that $Q_{A\rightarrow \tilde B} >0$, however 
the quantum information is not in all cases locally accessible.

The capability of a channel ${\cal E}$ 
to generate (distillable) pure state entanglement under LOCC$^{C}$ is 
completely determined by the entanglement properties of the state $E_{AB}$ \cite{Ci00}
corresponding to ${\cal E}$ via the isomorphism
\be 
E\equiv \eins_A\otimes{\cal E}_B |\Phi\rangle\langle \Phi|,\label{Iso}
\ee
and $|\Phi\rangle = 1/\sqrt{d}\sum_{k=1}^d |k\rangle_A|k\rangle_B$. We have 
that the 
channel can generate distillable pure state entanglement in the sense $A\tilde B$ iff 
$E$ is distillable (assuming LOCC$^C$ in both cases). To see this, assume
 on the one hand that $E$ is distillable. Since $E$ can be created by sending one half 
of a maximally entangled state through the channel ${\cal E}$, the channel 
can generate states which are distillable and thus pure entangled states. 
Assume on the other hand that by sending a certain state, say $\tau_A$, through the channel
 ${\cal E}$ one can generate distillable states $\sigma_{A\tilde B}$. Since $E$ can be 
used to implement ${\cal E}$ probabilistically by using only local operations and forward 
classical communication \cite{Ci00}, one can generate $\sigma_{A \tilde B}$ from $\tau_A$
 and $E$. Since $\sigma_{A \tilde B}$ is by assumption distillable, so is $E$.

Thus for both forward one way and two way classical communication and any subset of 
parties $\tilde B \in B$ we have the following result  \cite{note}:
\be 
Q_{A\rightarrow \tilde B}({\cal E})>0 \Leftrightarrow D_{A\tilde B}(E)>0
%Q_{\alpha}({\cal E})>0 &\Leftrightarrow& D_{\alpha}(E)>0 
\label{Q-D}
\ee
Note that for two--way classical communication Eq. (\ref{Q-D}) together with Eq. 
(\ref{equivD}) implies that if $D^\leftrightarrow_{AB_j}(E)=0 \forall B_j \in B$ 
then also all quantum capacities $Q^\leftrightarrow_{A\tilde B}$ are zero.

{\bf 4. Example for non--additivity of $Q_{A\rightarrow \tilde B}$}\\
We now use this fact to provide an example which shows that multipartite quantum 
channel capacities are not additive. In the following we restrict ourselves to 
two--way classical communication and thus omit the symbol $\leftrightarrow$. 
We consider a three party communication scenario and introduce three quantum 
channels ${\cal E}_a$, $a=1,2,3$ from a sender $A$ to two receivers $B$ and 
$C$ with ${\cal H}_c ={\cal H}_A =\C^4, {\cal H}_o={\cal H}_B \otimes 
{\cal H}_C =\C^2\otimes \C^2$. From now on we use the shorthand notation 
$Q$ to refer to any of the capacities $Q_{A\rightarrow B_1}, Q_{A\rightarrow B_2},
 Q_{A\rightarrow B_1B_2}$, since the results we obtain hold for any of these 
capacities. On the one hand, we show that each of the channels ${\cal E}_a$ 
is not capable to create pure state entanglement shared between any two of the
 parties $A,B,C$. It follows that the quantum capacities $Q$ of each channel 
${\cal E}_a$ are zero. On the other hand, we show that $Q$ of a channel 
$\bar{\cal E}$ created by randomly choosing one of the channels ${\cal E}_a$ 
with equal probability is non--zero which implies non--additivity of quantum 
channel capacities in this multi--party communication scenario. Note that quite
 remarkably the entanglement capability of the channels is enhanced by {\it classical 
mixing}, a procedure usually believed to diminish the entanglement properties of 
states and operations.

The quantum channels ${\cal E}_a$, $a=1,2,3$ are described by trace preserving 
completely positive maps which we write in the Kraus representation
%\be
$
{\cal E}\rho \equiv \sum_k A_k \rho A_k^\dagger.
$
%\ee
The map ${\cal E}_1$ is specified by the Kraus operators
$A_k \in$ $\{\frac{1}{4}(\sigma_0\sigma_0+\sigma_3\sigma_3)$, 
$\frac{1}{\sqrt{32}}(\sigma_1\sigma_1+\sigma_2\sigma_2)$, 
$\frac{1}{\sqrt{32}}(\sigma_2\sigma_1-\sigma_1\sigma_2)$,
$\frac{1}{4}\sigma_0\sigma_0$,
$\frac{1}{4}\sigma_3\sigma_3$,
$\frac{1}{4}\sigma_1\sigma_0$,
$\frac{1}{4}\sigma_2\sigma_0$,
$\frac{1}{4}\sigma_3\sigma_0$,
$\frac{1}{4}\sigma_0\sigma_1$,
$\frac{1}{4}\sigma_0\sigma_2$,
$\frac{1}{4}\sigma_0\sigma_3$,
$\frac{1}{4}\sigma_3\sigma_1$,
$\frac{1}{4}\sigma_1\sigma_3$,
$\frac{1}{4}\sigma_3\sigma_2$,
$\frac{1}{4}\sigma_2\sigma_3\}$
where $\sigma_j$ are the Pauli matrixes with $\sigma_0\equiv \eins_2$.
One readily checks that $\sum_k A_k^\dagger A_k = \eins$ which ensures that 
${\cal E}_1$ is trace preserving. 

Similarly, the map ${\cal E}_2$ is is determined by Kraus operators $A_k\in$
$\{\frac{1}{4}(\sigma_0\sigma_0+\sigma_3\sigma_3)$, 
$\frac{1}{4}[\sigma^-(\sigma_0+\sigma_3)]$, 
$\frac{1}{4}[\sigma^+(\sigma_0-\sigma_3)]$,
$\frac{1}{4}\sigma_0\sigma_0$,
$\frac{1}{4}\sigma_3\sigma_3$,
$\frac{1}{4}\sigma_3\sigma_0$,
$\frac{1}{4}\sigma_0\sigma_1$,
$\frac{1}{4}\sigma_1\sigma_1$,
$\frac{1}{4}\sigma_2\sigma_1$,
$\frac{1}{4}\sigma_3\sigma_1$,
$\frac{1}{4}\sigma_0\sigma_2$,
$\frac{1}{4}\sigma_1\sigma_2$,
$\frac{1}{4}\sigma_2\sigma_2$,
$\frac{1}{4}\sigma_3\sigma_2$,
$\frac{1}{4}\sigma_0\sigma_3\}$
where $\sigma^\pm \equiv(\sigma_1\mp i\sigma_2)/2$. Again one can easily
check that ${\cal E}_2$ is trace preserving. Finally, the map ${\cal E}_3$ 
is obtained from ${\cal E}_2$ by permuting systems 1 and 2.

To determine the (distillability) properties of the maps ${\cal E}_a$ we make 
use of the isomorphism 
%between completely positive maps and mixed states 
Eq. (\ref{Iso}). The mixed state $E_a$ associated to ${\cal E}_a$ 
is obtained by applying ${\cal E}_a$ to particles $B$,$C$ of the state 
$P_{\Phi}^{(A_1BA_2C)}\equiv|\Phi\rangle\langle\Phi|$ where $|\Phi\rangle\equiv |\Phi^+
\rangle_{A_1B}\otimes|\Phi^+\rangle_{A_2C}$ with $|\Phi^+\rangle=1/\sqrt{2}(|00\rangle+
|11\rangle)$. This corresponds to sending one part of a maximally entangled four level 
system through the quantum channel ${\cal E}$ from $A$ to $B$ and $C$ where qubits 
$A_1,A_2$ remain at site $A$. One finds that
\bea
E_1^{(A_1BA_2C)}=(2 |\Psi_0^+\rangle\langle\Psi_0^+| + \eins - P_{1010}-P_{0101})/16 \nonumber\\
E_2^{(A_1BA_2C)}=(2 |\Psi_0^+\rangle\langle\Psi_0^+| + \eins - P_{0100}-P_{1011})/16, 
\eea
where $P_{ijkl}\equiv |ijkl\rangle\langle ijkl|$ and $|\Psi_0^+\rangle\equiv 
1/\sqrt{2}(|0000\rangle+|1111\rangle)$ and $E_3$ is obtained from $E_2$ by
 exchanging $(A_1B)$ with $(A_2C)$. One can readily determine the 
entanglement properties of these density operators. We have that 
$E_1^{T_{B}}\geq 0, E_1^{T_{C}}\geq 0$ and $E_2^{T_{A_1A_2}} \geq 0, E_2^{T_{C}}\geq 0$ 
and similarly for $E_3$, where we denote by $E^{T_\beta}$ the partial transposition of 
the density operator $E$ with respect to party $\beta$ \cite{Pe96,Du00}. Note that $E_a$
 belongs to a class of states which has been completely characterized with
respect to its entanglement properties in Ref. \cite{Du00}: for those states
positivity of partial transposition with respect to a certain system already 
implies separability of this system from the remaining ones, e.g. $E_2 = \sum_k p_k |\phi_k\rangle_{C}\langle\phi_k| \otimes|\chi_k\rangle_{A_1BA_2}\langle\chi_k|$ follows from $E_2^{T_C} \geq 0$. 
It follows that the channel ${\cal E}_1$ can only create states where particle $B$ 
(and also particle $C$) is separable and thus no pure state entanglement shared between
 any of the groups $(A_1A_2), B, C$ can be created, i.e. $D_{AB}^\leftrightarrow(E_1)= 
D_{AC}^\leftrightarrow(E_1)=0$ \cite{Du00}. 
%This is due to the fact that the partial transpose e.g. with respect to system $B$ remains positive under any sequence of local operations $(A_1A_2)- B -C$ while it has to become non--positive if a pure entangled state share between parties $(A_1A_2)-B$ should be created. 
We thus have that the quantum capacity
 of ${\cal E}_1$ is zero, $Q({\cal E}_1)=0$. Similarly, we have that $E_2, E_3$ are 
separable with respect to system $(A_1A_2)$ which leads to $Q({\cal E}_2) = Q({\cal E}_3)
 =0$.
Note that the considered channels are multipartite binding entanglement 
channels since they are isomorphic to bound entangled states $E_{a}$.

We consider now the situation where all three channels ${\cal E}_a$ are available 
and show that $Q({\cal E}_1,{\cal E}_2,{\cal E}_3) > 0$. In particular, we consider
 a channel $\bar{\cal E}$ which is obtained by a classical mixture of the three 
channels ${\cal E}_a$, i.e. randomly choosing one of the channels. The new set of 
Kraus operators for the channel $\bar{\cal E}$ is given by the Kraus operators of 
the channels ${\cal E}_a$ with a pre-factor $1/\sqrt{3}$. The density operator $\bar E$ 
corresponding to  $\bar{\cal E}$ is given by $\bar E =(E_1+E_2+E_3)/3$ and one finds
\bea
&&\bar E^{(A_1BA_2C)}=1/16(2 |\Psi_0^+\rangle\langle\Psi_0^+| + \eins - \\
&&- 1/3[P_{1010}+P_{0101}+P_{0100}+P_{1011}+P_{0001}+P_{1110}]).\nonumber
\eea
We have $\bar E^{T_{A_1A_2}}\not\geq 0, 
 \bar E^{T_{B}}\not\geq 0, \bar E^{T_{C}}\not\geq 0$ which implies that $\bar E$ 
is distillable entangled, i.e. maximally entangled pure states shared between 
$(A_1A_2)-B$ and $(A_1A_2)-C$ can be created. This is due to the fact that $\bar E$ 
belongs to a class of states $\rho$ for which non--positive 
partial transposition with respect to all relevant partitions is a sufficient condition 
for distillability \cite{Du00}. Note that the distillation procedure requires two--way classical 
communication. From Eq. (\ref{Q-D}) follows that also the quantum capacity $Q(\bar{\cal E}) 
> 0$, which implies that the quantum capacity of multiparty communication channels is not 
additive.

In this paper, we have considered multiparty communication scenarios where information 
is sent from a single sender to several receivers. We have introduced (natural) definitions
 of quantum channel capacities in the multipartite setting and established a connection of 
these quantum capacities to the capability of the channel to create distillable entangled 
states. This connection allowed us to show that all these quantum capacities are --in the 
case of two way classical communication-- not additive.

 %\acknowledgements
W.D. thanks H.-J- Briegel and F. Verstraete for discussions. This work was supported 
by European Union (IST-2001-37559,-38877,-39227), the DFG and the 
Austrian Academy of Science (APART  -W.D.).

 %----------------------------------------------------------------------------------------------------- 
 %-----------------------------------------------------------------------------------------------------

 %\begin{references}

 %\end{references}

\begin{thebibliography}{99}


\bibitem{Sh48}
E. Shannon, Bell Syst. Tech. J. {\bf 27}, 379 (1948).

\bibitem{Sc95}
B. Schumacher, Phys. Rev. A {\bf 51}, 2738 (1995).

\bibitem{Jo94}
R. Jozsa and B. Schumacher, J. Mod. Opt. {\bf 41}, 2343 (1994);
%/bibitem{Ba96}
H. Barnum {\em  et al.}, 
%, C.A. Fuchs, R. Jozsa, and B. Schumacher, Phys. Rev. A
Phys. Rev. A {\bf 54}, 4707 (1996); 
R. Jozsa {\em et al.},
%, P. Horodecki, M. Horodecki, and R. Horodecki, 
Phys. Rev. Lett {\bf 81}, 1714 (1998).

\bibitem{Shor03}
P. Shor, lecture notes;  
%from MSRI Workshop on quantum computation, 2002, avalable at http://www.msri.org/publications/ln/msri/2002/
%quantumcrypto/shor/1/; 
I. Devetak, quant-ph/0304127.

\bibitem{Be97}
C. H. Bennett {\em et al.}, 
%, D. P. Di Vincenzo, J. Smolin and W. K. Wootters, 
Phys. Rev. A {\bf 54}, 3814 (1997).

\bibitem{bound}
M. Horodecki {\em et al.}, Phys. Rev. Lett. {\bf 80}, 5239 (1998).

%\bibitem{activ}P. Horodecki {\em et al.}, Phys. Rev. Lett. {\bf 82}, 1056 (1999).

\bibitem{bechan}
P. Horodecki {\em et al.}, J. Mod. Opt.  {\bf 47}, 347 (2000).

\bibitem{APP} For this subject see [P. Horodecki, Acta Physica Polonica A, 
{\bf 103}, 399 (2002)] and references therein.

\bibitem{CMP}
D. P. DiVincenzo {\em et al.}, Comm. Math. Phys.  {\bf 238}, 379 (2003).

\bibitem{CEJP}
%P. Horodecki, Central European Journal of Physics {\bf 4}, 695 (2003).
P. Horodecki, CEJP {\bf 4}, 695 (2003).


\bibitem{Ba98}
H. Barnum, M. Knill and M. A. Nielsen, 
IEEE Trans.Info.Theor. {\bf 46}, 1317 (2000). (quant-ph/9809010).

\bibitem{Notefeatures}
One may e.g. establish relations between different kind of capacities, such 
as between $Q_{\tilde A \rightarrow B_j}$ and $Q_{\tilde A \tilde B}$ with 
$B_j \in \tilde B$. One finds $\frac{1}{M} \sum_{j=1}^M Q_{\tilde A \rightarrow B_j} 
\leq Q_{\tilde A \rightarrow \tilde B},\label{relQ}$ which can be seen by noting that 
$Q_{\tilde A \rightarrow B_j} \leq Q_{\tilde A \rightarrow \tilde B}$, where equality
 in this expression can be {\em simultaneously} achieved $\forall j$ (consider e.g. a
 channel which maps $|j\rangle \rightarrow |j\rangle^{\otimes N}$, $j=0,1$, which has 
$Q_{\tilde A \rightarrow \tilde B}^\leftrightarrow= Q_{\tilde 
A \rightarrow B_j}^\leftrightarrow =1$). Note, however, that this does not imply
 that quantum information can be sent at this rate simultaneously to all receivers.

\bibitem{Po92}
S. Popescu and D. Rohrlich, Phys. Lett. A. {\bf 166}, 293 (1992).

%\bibitem{Ja72} 
%A. Jamiolkowski, Rep. Math. Phys. {\bf 3}, 275, (1972).

\bibitem{Ci00}
J. I. Cirac {\em et al.}, 
%, W. D\"ur, B. Kraus and M. Lewenstein, 
Phys. Rev. Lett. {\bf 86}, 544 (2001).


\bibitem{note}
Note that one can prove $Q_{A\rightarrow \tilde B}({\cal E}) > 0 
\Rightarrow D_{A\tilde B}(E)>0$ also in a more formal way following 
closely the proof given in Ref. \cite{CEJP} for the case of bipartite channels.


\bibitem{Pe96}
A. Peres, Phys. Rev. Lett. {\bf 77}, 1413 (1996).

\bibitem{Du00}
W. D\"ur and J. I. Cirac, Phys. Rev. A {\bf 61}, 042314 (2000); %classification of multiqubit mixed states
W. D\"ur and J. I. Cirac, Phys. Rev. A {\bf 62}, 022302 (2000). %Activation of bound entanglement



%\bibitem{gen}
%M. Horodecki, P. Horodecki and R. Horodecki,
%Phys. Rev. A {\bf 60},  1888 (1999).

\end{thebibliography}
\end{document}